\documentclass{aastex62}
\usepackage[latin9]{inputenc}
\usepackage{amsbsy}
\usepackage{graphicx}

\makeatletter
\pdfoutput=1

\makeatother

\submitjournal{ApJL}

\begin{document}

\title{Evolution of Magnetic Helicity in Solar Cycle 24}

\correspondingauthor{Valery V. Pipin}
\email{pip@iszf.irk.ru}

\author[0000-0001-9884-1147]{Valery V. Pipin}
\affil{Institute for Solar-Terrestrial Physics \\
PO Box 291, Lermontov st., 126a \\
Irkutsk, 664033, Russia}

\author[0000-0003-0489-0920]{Alexei A. Pevtsov}
\affil{National Solar Observatory \\
3665 Discovery Drive, 3rd Floor \\
Boulder, CO 80303, USA}
\affil{Central Astronomical Observatory
of RAS at Pulkovo \\
Saint Petersburg, 196140, Russia}

\author{Yang Liu}
\affil{Stanford University, Stanford, CA, USA}

\author[0000-0003-0364-4883]{Alexander G. Kosovichev}
\affil{New Jersey Institute of Technology, Newark, NJ USA}

\begin{abstract}
We propose a novel approach to reconstruct the surface magnetic helicity density on the Sun or sun-like stars. 
The magnetic vector potential is determined via decomposition of vector magnetic field measurements into
toroidal and poloidal components. The method is verified using data from a non-axisymmetric dynamo 
model. We apply the method to vector field synoptic maps from Helioseismic and Magnetic Imager (HMI) on board of Solar Dynamics Observatory (SDO) to study evolution of the magnetic
helicity density during solar cycle 24. It is found that the mean helicity density of the non-axisymmetric magnetic field of the Sun evolves in a way which is similar to that reported for the current helicity density of the solar active regions. It has predominantly the negative sign in the northern hemisphere, and it is positive in the southern hemisphere. Also,  the hemispheric helicity rule for the non-axisymmetric magnetic field showed the sign inversion at the end of cycle 24.  Evolution of magnetic helicity density of large-scale axisymmetric magnetic field is different from that expected in dynamo theory. On one hand, the mean large- and small-scale components of magnetic helicity density display the hemispheric helicity rule of opposite sign at the beginning of cycle 24. However, later in the cycle, the two helicities exhibit the same sign in contrast with the theoretical expectations.
\end{abstract}


\section{Introduction}
\label{sec:intro}

Magnetic helicity is an integral measure of topological properties of the magnetic
field in closed volume $\rm V$:
\begin{equation}
H_{M}=\int\mathbf{A}\cdot\mathbf{B}d\mathrm{V},\label{eq:def}
\end{equation}
where $\mathbf{A}$ is the magnetic vector potential, $\mathbf{B}=\boldsymbol{\nabla}\times\mathbf{A}$,
and the $\mathbf{B}$ is confined to the volume $\mathrm{V}$.
Locally, it can be characterized by a number of parameters such
as linkage, twist and writhe of the field lines \citep{2018JPhA51W5501B}.
In astrophysical dynamos, magnetic helicity is commonly accounted as a nonlinear constraint
of turbulent generation of large-scale magnetic field \citep{pouquet-al:1975b,kleruz82,brsu05}. 
Computation of magnetic helicity on the Sun requires knowledge of vector magnetic field in a 3D region, but observations are usually taken 
in a shallow layer of the solar atmosphere (typically, in the photosphere). Thus, early studies turned to calculation of so-called helicity proxies, such as, for example, vertical components of current helicity density $J_z \cdot B_z$ or $\alpha = J_z / B_z$ (a measure of magnetic twist).
\citet{see1990SoPh} and \citet{pev95ApJ} found that
the current helicity density and twist in solar active regions follow $\it {the~ hemispheric~
helicity~ rule}$ with predominantly negative values in the northern hemisphere and positive values in the southern hemisphere. The hemispheric preference for
the helicity sign was later confirmed by several researchers \citep[e.g.,][]{long1998ApJ,baoZh98,hagsak05}.
\citet{choud2004ApJ} proposed that due to interaction of the large-scale toroidal field from a previous solar cycle and the poloidal field of a new cycle, the hemispheric helicity could reverse sign at the beginning of each cycle. \citet{hagsak05} 
reported the presence of such sign reversal in 
the Okayama Observatory Solar Telescope (OAO) and the Mitaka Solar Flare Telescope (SFT) observations. 
\citet{Pevtsov.etal2008} examined the periods of sign reversals observed by different instruments and found no agreement among the datasets during these periods.
They concluded that at least some reversals could be due to statistical nature of the hemispheric helicity rule. 
Later, \citet{zetal10} showed that hemispheric helicity rule of the
solar active regions evolve with time and found its reversals at the beginning of Solar Cycles 22 and 23. 
\citet{sok2013} found inversions of the current helicity density in active regions at the 
beginning and at the end of the solar cycle 23.
These early findings are
based on vector magnetograms of active regions only.
Systematic full disk observations of vector magnetic 
field became available in 2009 from Vector Spectromagnetograph (VSM) on Synoptic Optical Long-term Investigations of the Sun (SOLIS) platform \citep{Balasubramaniam.Pevtsov2011} and in 2010 from
Helioseismic and Magnetic Imager \citep[HMI, ][]{Scherrer.etal2012} on board Solar Dynamics 
Observatory \citep[SDO,][]{Pesnell.etal2012}. Prior to these observations, calculations of current 
helicity relied on vector magnetic field 
reconstructed from rotational modulation of the 
observed longitudinal (line-of-sight) field 
\citep{2000ApJ...528..999P}. 
Later studies demonstrated that the sign of 
helicity of large-scale magnetic fields is opposite to the sign of helicity of active regions 
\citep{PP14,2017ApJ836.21B}. The large-scale 
helicity was also found to evolve during solar 
cycles similarly to helicity of active regions. 

The solar dynamo theory predicts bi-helical properties of magnetic fields \citep{black-bran:02,brsu05}. In this theory the sign of magnetic helicity density of large-scale field 
should correspond to the sign of the $\alpha$-effect, while sign of magnetic helicity density of the small-scale field would result from the magnetic helicity conservation, and thus, it would be opposite to large-scale helicity. In the framework of this model,  small-scale magnetic field corresponds to active regions and the large-scale stands for the global axisymmetric components of the solar magnetic activity.  The bi-helical properties were studied recently by \cite{2017ApJ836.21B} and \cite{2018arXiv180404994S} using
two-scale approximation and the vector magnetic 
field measurements from  SDO/HMI and  SOLIS. The 
results from two different instruments appear to 
be  inconclusive in respect to bi-helical nature of solar magnetic field.

Other predictions of mean-field dynamo models
include the existence of polar and equatorial branches in the time-latitude diagram of magnetic helicity evolution. According to \citet{pip13M}, those branches  represent
the transport of magnetic helicity flux to the polar regions, both on large and small scales.
In this paper we present the first observational evidence of polar branches of the large-scale magnetic helicity in solar cycle 24. 
Section \ref{sec:method} describes our method of calculation of magnetic helicity. Section \ref{sec:bench} verifies the
proposed methodology using synthetic data from mean-field dynamo model calculations. 
Section \ref{sec:obs_helicity} presents the derivation of 
magnetic helicity using observed magnetic fields, 
and Section \ref{sec:discussion} discusses our findings.

\section{The method}
\label{sec:method}

To determine the magnetic helicity density, we employ a decomposition of the vector magnetic field
into toroidal and poloidal components using scalar potentials
$S$ and $T$ \citep{KR80,2018JPhA51W5501B}:
\begin{eqnarray}
\mathbf{B} & = & \boldsymbol{\nabla}\times\left(\hat{\mathbf{r}}T\right)+\boldsymbol{\nabla}\times\boldsymbol{\nabla}\times\left(\hat{\mathbf{r}}S\right) = \nonumber \\
 & = & -\frac{\hat{r}}{r}\Delta_{\Omega}S+\hat{\theta}\left(\frac{1}{\sin\theta}\frac{\partial T}{\partial\phi}-\frac{\sin\theta}{r}\frac{\partial F_{S}}{\partial\mu}\right)+\hat{\phi}\left(\sin\theta\frac{\partial T}{\partial\mu}+\frac{1}{r\sin\theta}\frac{\partial F_{S}}{\partial\phi}\right)
 \label{eq:pol_tor}
\end{eqnarray}
where ${\displaystyle \Delta_{\Omega}=\frac{\partial}{\partial\mu}\sin^{2}\theta\frac{\partial}{\partial\mu}+\frac{1}{\sin^{2}\theta}\frac{\partial^{2}}{\partial\phi^{2}}}$,
$\mu=\cos\theta$ and $\theta$ is the polar angle, and $F_{S}=\partial\left(rS\right)/\partial r$.
Three components of vector magnetic field
\citep[radial $r$, meridional $\theta$, and zonal $\phi$, see,][]{Virtanen.etal2019}
are then represented by three independent variables,
$S$ , $T$, and $F_{S}$:
\begin{eqnarray}
B_{r} & = & -\frac{1}{r}\Delta_{\Omega}S,\label{eq:S}\\
B_{\theta} & = & \frac{1}{\sin\theta}\frac{\partial T}{\partial\phi}-\frac{\sin\theta}{r}\frac{\partial F_{S}}{\partial\mu},\label{eq:fs}\\
B_{\phi} & = & \sin\theta\frac{\partial T}{\partial\mu}+\frac{1}{r\sin\theta}\frac{\partial F_{S}}{\partial\phi}\label{eq:T}
\end{eqnarray}
To determine a unique solution of Equations \ref{eq:S}--\ref{eq:T} we apply the following gauge
(see, e.g., \citealt{KR80}): 
\begin{equation}
\int_{0}^{2\pi}\int_{-1}^{1}Sd\mu d\phi=\int_{0}^{2\pi}\int_{-1}^{1}Td\mu d\phi=\int_{0}^{2\pi}\int_{-1}^{1}F_{S}d\mu d\phi=0.\label{eq:norm}
\end{equation}
Note, that in the case of potential magnetic field $T=0$, and $F_{S}$
is determined by $S$. Therefore, the system of Equations \ref{eq:S}--\ref{eq:T} represents a least-squares 
problem. 
Hereafter we consider the general case
of nonpotential magnetic fields on the solar surface. Eqs.(\ref{eq:S}), (\ref{eq:fs}) and (\ref{eq:T}) can be transformed
in:
\begin{eqnarray}
-\frac{1}{r}\Delta_{\Omega}S & = & B_{r}\\
\Delta_{\Omega}T & = & \frac{\partial}{\partial\mu}\sin\theta B_{\phi}+\frac{1}{\sin\theta}\frac{\partial B_{\theta}}{\partial\phi},\\
\frac{1}{r}\Delta_{\Omega}F_{S} & = & \frac{1}{\sin\theta}\frac{\partial B_{\phi}}{\partial\phi}-\frac{\partial}{\partial\mu}\sin\theta B_{\theta}
\end{eqnarray}
Reconstruction and differentiating is done in the spectral spherical harmonic space using
the SHTools \citep{sht2018}. After finding solutions for  $S$, $T$ and  $F_{S}$
we can determine components of vector potential $\mathbf{A}$, 
\begin{eqnarray}
\mathbf{A} & = & \hat{\mathbf{r}}T+\boldsymbol{\nabla}\times\left(\hat{\mathbf{r}}S\right) = \nonumber \\
 & = & \hat{\mathbf{r}}T+\frac{\hat{\theta}}{\sin\theta}\frac{\partial S}{\partial\phi}+\hat{\phi}\frac{\sin\theta}{r}\frac{\partial S}{\partial\mu}.
\end{eqnarray} 
Below we demonstrate the method using output of a dynamo model, where
we have complete information about the distribution of vector magnetic field, its vector potential and
magnetic helicity density.

\section{Dynamo Model Benchmark}
\label{sec:bench}

We use results of  the non-axisymmetric $\alpha^{2}\Omega$
dynamo model developed recently by \citet{2018ApJ867.145P}. The model
simulates the solar-type dynamo, in which the surface magnetic activity is governed by the dynamo-generated axisymmetric toroidal magnetic field. 
The dynamo parameters of the model are below threshold of the dynamo instability 
of large-scale non-axisymmetric magnetic fields. 
This resembles the situation for solar-type dynamos
\citep{rad86AN}. In order to mimic emergence of solar active regions,
we take into account the Parker's magnetic buoyancy instability, which
produces bipolar regions from the toroidal magnetic field at random
latitudes and random moment of time under the condition that the magnetic
field strength exceeds a critical threshold. The reader can find a detailed description
of the model and the code in \citet{2018ApJ867.145P} and \citet{2dspy}.

\begin{figure}
\includegraphics[width=1\textwidth]{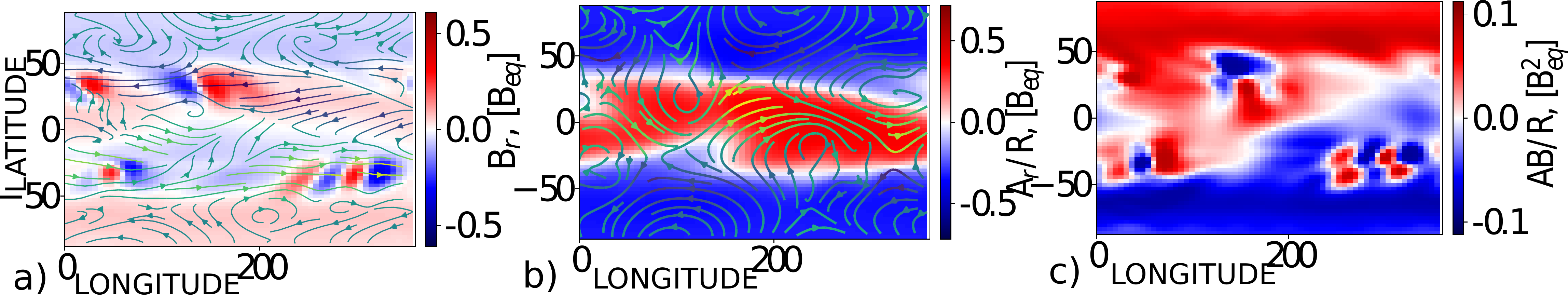}

\caption{\label{snap}a) Snapshot of the radial magnetic field distribution
(color image), streamlines show the horizontal (toroidal and poloidal)
magnetic field; b) snapshot for the magnetic helicity density;
c) the radial component of the current helicity density $H_{C}=B_{r}\left(\boldsymbol{\nabla}\times\mathbf{B}\right)_{r}$.}

\end{figure}

Figure \ref{snap} shows snapshots of the magnetic field components, as well as the magnetic and current helicity densities. The snapshots are taken during the maximum of the dynamo cycle. The magnetic helicity has predominantly positive sign in the northern hemisphere and the negative sign  in the southern hemisphere.
In this case, the hemispheric helicity rule of the large-scale magnetic field follows from the dynamo theory which predicts that the sign of the magnetic helicity of the large-scale field corresponds to the sign of the $\alpha$-effect in a given hemisphere \citep{bl-br2003,2018arXiv180404994S}. 

\section{Magnetic Helicity Density Derived From Observations}
\label{sec:obs_helicity}

\subsection{Observational Data}

We apply the formalism described in Section \ref{sec:method}
to a set of intermediate resolution (360 by 720 pixels) vector magnetic fields
synoptic maps from HMI/SDO. The dataset includes 116 
Carrington rotations (CR) from CR2097 (May 2010) to CR2214 
(March 2019). The HMI synoptic maps are calculated in Carrington longitude (degrees) -- sine (latitude) coordinate grid. The pixel size is 0.5 degree in longitude and 1/180 in sine latitude. Method for producing the HMI synoptic maps is described in details by \cite{2017SoPh292.29L}.
For these data, the 180 degree ambiguity in the horizontal field direction
was resolved by the HMI team using a combination of 
a minimum energy criterium (for pixels with stronger fields) and 
random disambiguation (for weak field pixels). Additional 
details about the HMI data reduction and the disambiguation procedure 
can be found in the above cited paper.

\cite{2017SoPh292.29L} found  that with the chosen combination of disambiguation 
methods   the noise level of the vector magnetic field in the synoptic charts 
varies from $\pm 10$G during the solar maximum to  $\pm 20$G during the solar minimum. 

\subsection{Results}

\begin{figure}
\begin{centering}
\includegraphics[width=0.95\textwidth]{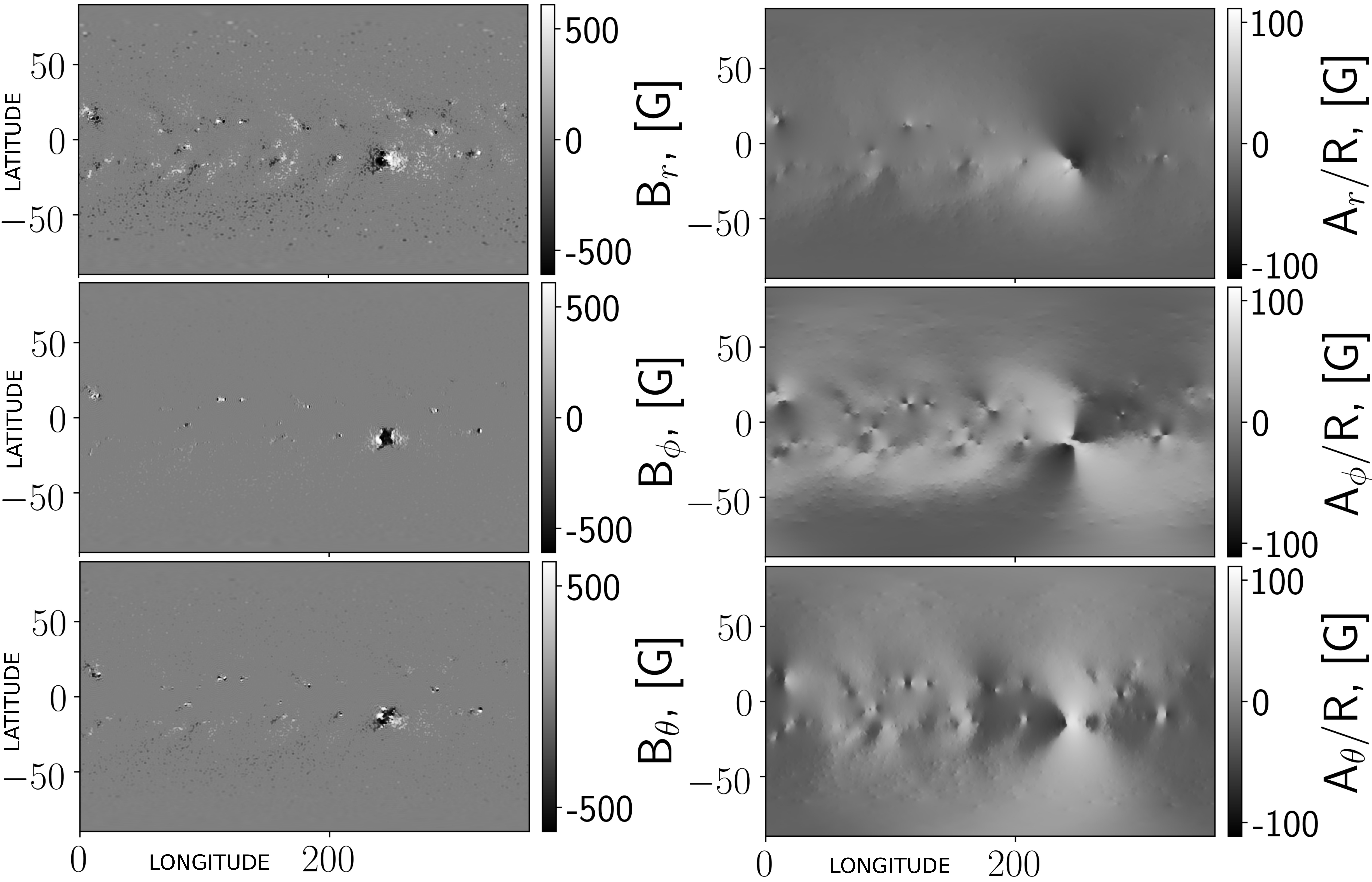}
\par\end{centering}
\caption{\label{synm}Synoptic maps of the vector magnetic field and the reconstructed
potential from SDO/HMI for CR 2156.}
\end{figure}

\begin{figure}
\begin{centering}
\includegraphics[width=0.5\textwidth]{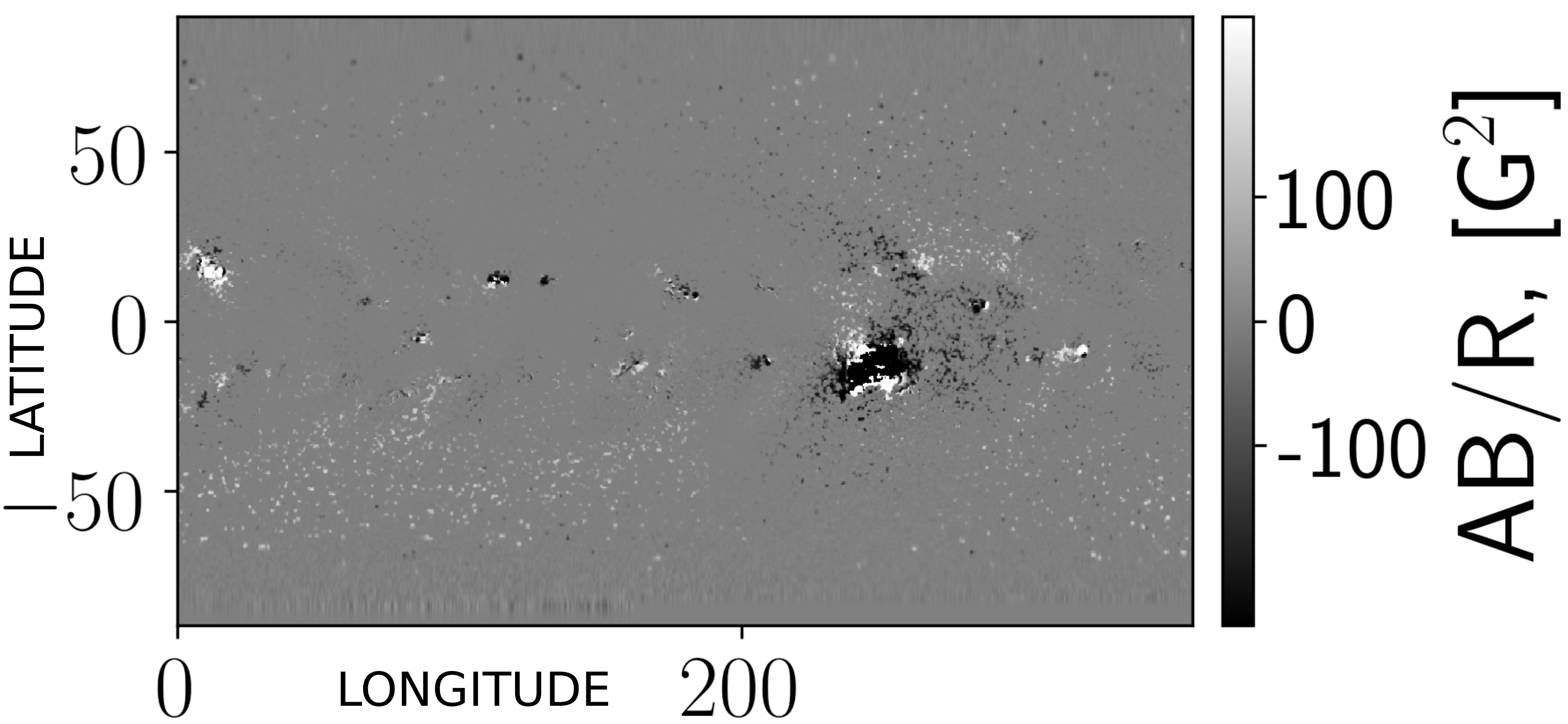}
\par\end{centering}
\caption{\label{recon}Distribution of the magnetic helicity density $\mathbf{A}\cdot\mathbf{B}$  for CR 2156.}
\label{fig:cr2156}
\end{figure}

As the first example, we consider results of reconstruction of the vector potential for the synoptic maps of CR2156. This Carrington rotation is characterized by strong active region NOAA AR~12192 that emerged in the southern hemisphere. Figure \ref{synm}
shows  the synoptic maps of the vector magnetic field and the reconstructed
potentials $\mathrm{S}$, $\mathrm{A_{r}=rT}$. Distribution of the vector potential reveals a large-scale non-axisymmetric pattern associated with this active region. Interesting that the vector-potential components show inverse sign relative to the corresponding magnetic field components in the  core of  the active region. This  results in  predominantly negative magnetic helicity  in AR~12192, which is shown in Figure \ref{recon}. We notice that the east part of the southern hemisphere shows the background magnetic helicity of the positive sign which corresponds to the basic hemispheric helicity of the active regions \citep{Pevtsov.etal2014}.  The large area of the negative magnetic helicity density around  NOAA AR~12192 resembles the situation demonstrated in our dynamo model. Thus, it can be speculated that the origin of this active region is related to large-scale magnetic field located in a shallow subsurface layer.  

\begin{figure}
\begin{centering}
\includegraphics[width=0.95\textwidth]{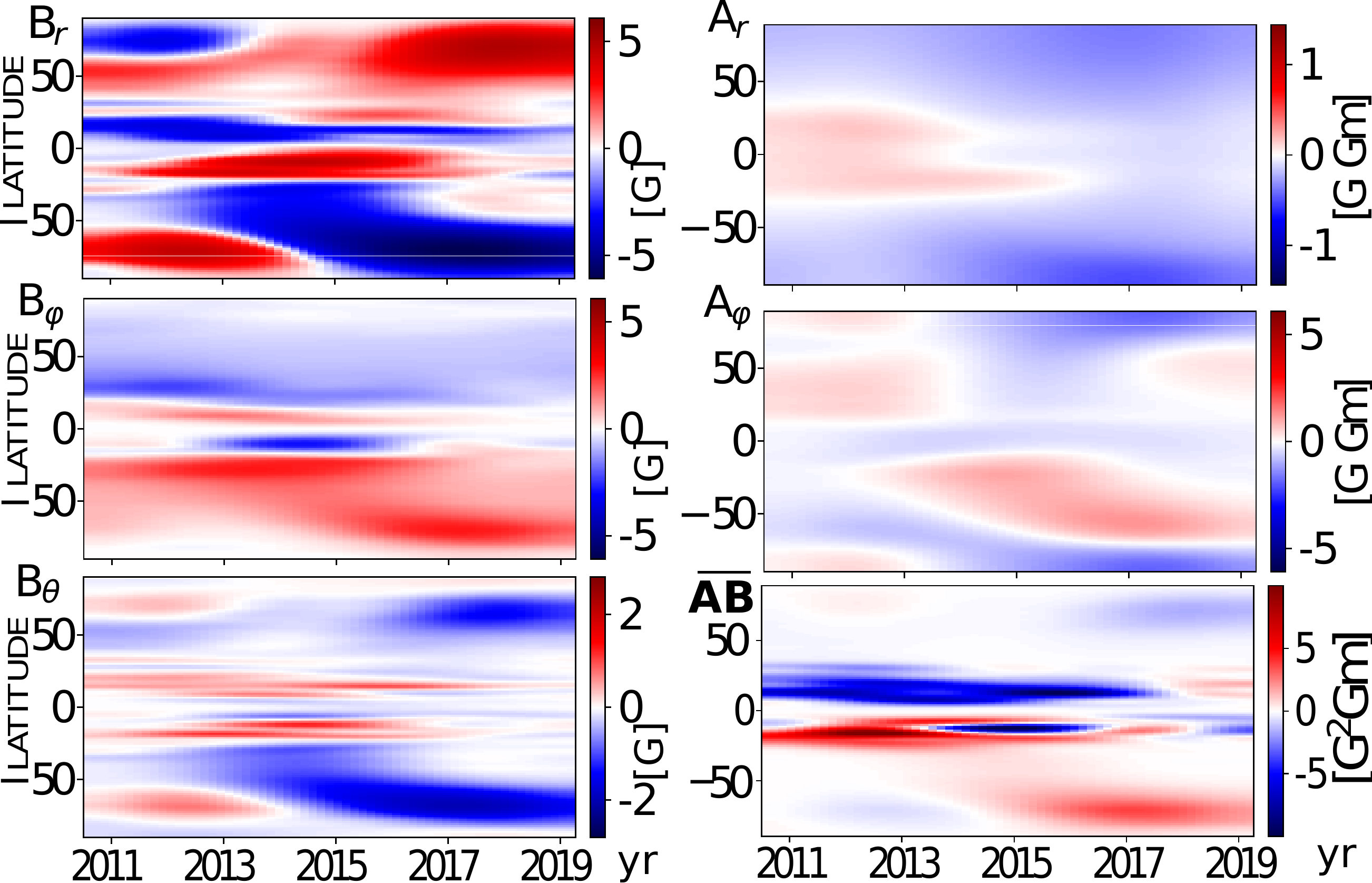}
\par\end{centering}
\caption{\label{bhmi}The time-latitude evolution of the large-scale magnetic
field : $\bar{B}_r$, $\bar{B}_{\phi}$, and $\bar{B}_{\theta}$, as well as the large-scale vector-potential components $\bar{A}_r$ and $\bar{A}_{\phi}$, the total magnetic helicity density $\overline{\mathbf{{A}\cdot\mathbf{{B}}}}$.}

\end{figure}

Solving  Equations \ref{eq:S}--\ref{eq:T} for each Carrington rotation from CR2097 to CR2214 we derived the distributions of the vector potential and magnetic helicity density for solar cycle 24. The large-scale distributions of magnetic field and helicity density are obtained by means of the azimuthal averaging of the synoptic maps for each Carrington rotation. To represent the mean signal, where we filtered out the time variations with period less than 2 year. Also, we employ the Gaussian smoothing with a FWHM equal 5 pixels in latitude and 10 CR time. Figure \ref{bhmi} shows the time-latitude diagrams for the axisymmetric components of magnetic field and vector potential. The time-latitude diagrams of the magnetic field evolution are in agreement with \cite{2018MNRAS480477V}. The reconstructed potentials disagree with results of \cite{PP14}, who used profiles of $\bar{B}_r$ from different longitudinal distances from the central meridian to reconstruct $\bar{B}_{\phi}$ and $\bar{A}_{r}$. Also, the obtained $\bar{B}_{\phi}$ disagrees with our results at the polar-ward side of the sunspot activity zone, where the large-scale toroidal magnetic field is present. The sign of this field is opposite to the sign of toroidal field in the solar active regions. Such bi-modal structure of the axisymmetric toroidal magnetic field affects the  distribution of $\bar{A}_{r}$. The sign of the polar-ward side of the axisymmetric toroidal magnetic field  can hardly be explained by surface effect of the differential rotation acting on the meridional component of the axisymmetric magnetic field. From Figure \ref{bhmi} it is clear that the direction of the meridional component is opposite to the one which might produce the polar-side toroidal magnetic field.  
Also, in our case, the evolution of $\bar{A}_{\phi}$ differs from results of \cite{sten88,bl-br2003} and \cite{PP14} as well, where they found symmetric $\bar{A}_{\phi}$ profiles relative to the equator. The difference is likely due to the rather asymmetric development of Cycle 24 in the northern and southern hemispheres.

\begin{figure}
\begin{centering}
\includegraphics[width=0.95\textwidth]{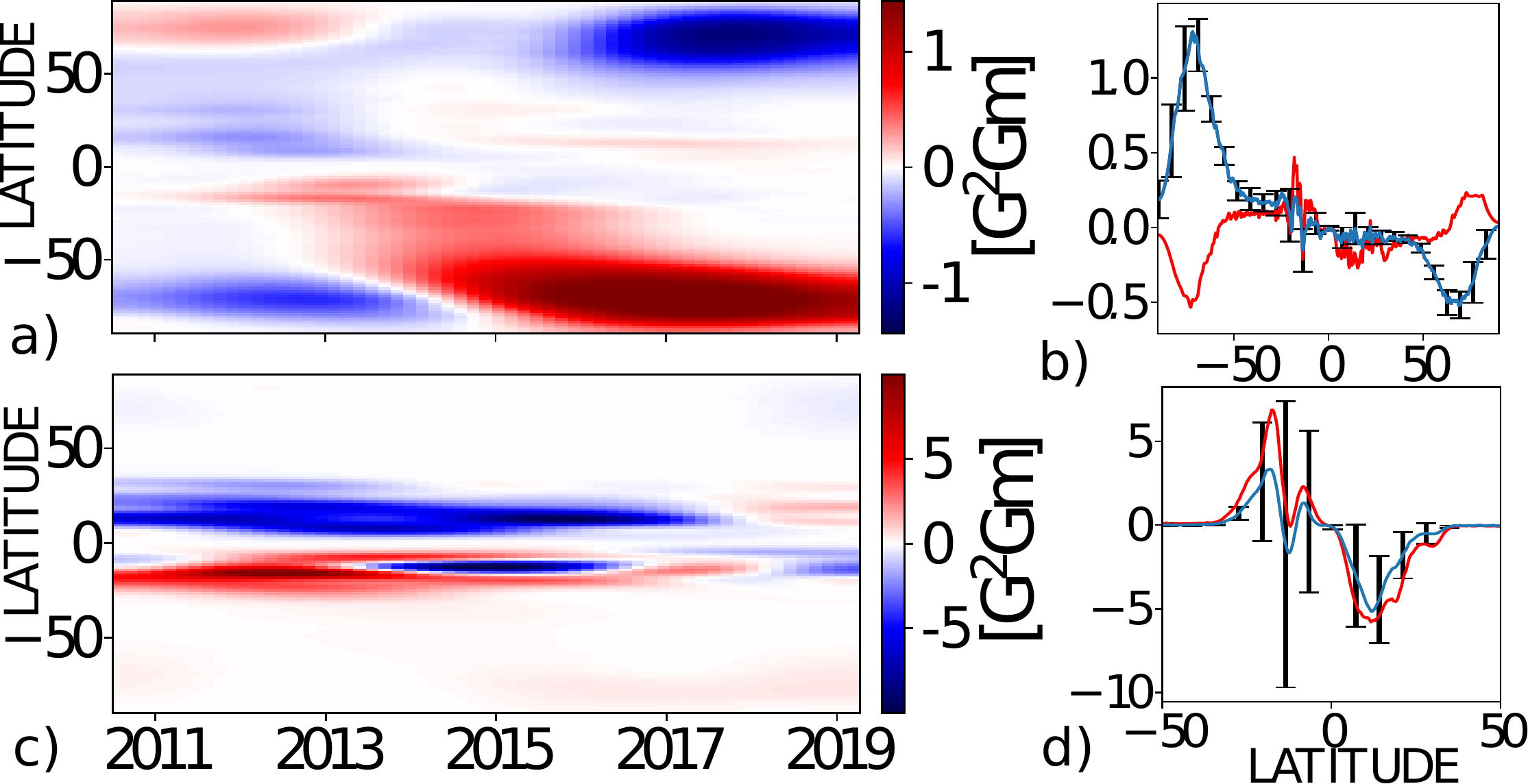}
\par\end{centering}
\caption{\label{hhmi} Magnetic helicity in Solar Cycle 24: a) The time-latitude evolution of the large-scale magnetic
helicity density, $\mathbf{\overline{A}\cdot\mathbf{\overline{B}}}$; b) the mean latitudinal profile of the large-scale magnetic helicity density with 95$\%$ confidence interval for the standard error, the red line shows the average profile for the first half of the cycle including CR2097-2156; c) shows the same as (a) for 
the azimuthal averaging of the small-scale magnetic helicity density
$\mathbf{\overline{a\cdot\mathbf{b}}}=\overline{\mathbf{{A}\cdot\mathbf{{B}}}}-
\mathbf{\overline{A}\cdot\mathbf{\overline{B}}}$; d) shows the same as (b) for the
$\mathbf{\overline{a\cdot\mathbf{b}}}$.}
\end{figure}

Figure \ref{hhmi}a shows the time-latitude diagrams of the magnetic helicity density of the large-scale field, $\bar{\mathbf{A}}\cdot\bar{\mathbf{B}}$. The magnetic helicity of the large-scale
field is positive in the high-latitude zone of the northern hemisphere at the beginning of Cycle 24, and it changes the sign to negative after the polar field reversal. A similar  situation is observed in the southern hemisphere. 
Figure \ref{hhmi}b shows the time-averaged hemispheric helicity sign rule together with uncertainty bars. The  uncertainties are estimated by using differences of the original and smoothed signals. It is seen that in average over the cycle 24 the large scale magnetic field shows the same  hemispheric sign as the  ``small-scale'' magnetic field. However if we would restrict the averaging period to the first half of the cycle 24, i.e., by CR2097--2156 we get, in general, the positive magnetic helicity density in the northern hemisphere and negative in the southern one (except the low latitudes). Therefore at this period of time the bi-helical property can be confirmed with some reservations. This agrees with the two-scale analysis  of \cite{2017ApJ836.21B} as well as with results of \cite{PP14} (cf., Fig9a there). 
Also, Figures \ref{hhmi}c and d support this conclusion.  

Figure \ref{hhmi}c  shows evolution of 
 ``small-scale'' magnetic helicity density, i.e., $\mathbf{\overline{a\cdot\mathbf{b}}}=\overline{\mathbf{{A}\cdot\mathbf{{B}}}}-
\mathbf{\overline{A}\cdot\mathbf{\overline{B}}}$.  The definition of small-scale helicity, $\mathbf{\overline{a\cdot\mathbf{b}}}$,  includes magnetic fields from all range of  scales except the axisymmetric magnetic field. A more accurate  analysis  of the magnetic helicity distribution over the scales can be done by using the two-scale analysis introduced by \citet{2017ApJ836.21B}.
Averaged  over the cycle  $\mathbf{\overline{a\cdot\mathbf{b}}}$, shows the hemispheric helicity rule for the solar active regions.  It shows the large uncertainty  bars  which are  mostly caused by fluctuations in magnetic flux emergence. It is also can be seen that for the first half of the cycle 24 the  $\mathbf{\overline{a\cdot\mathbf{b}}}$ holds  the hemispheric helicity rule. The violation of the rule in the time-averaged signal is likely caused by the magnetic activity in the Southern hemisphere during 2014--2015 years. In particular, the active region NOAA 12192 strongly violates the hemispheric helicity rule as we see in Figure \ref{fig:cr2156}. In the end of our observational period, the  $\mathbf{\overline{a\cdot\mathbf{b}}}$ shows inversion of helicity sign in the low latitudes. 
This agrees with the results for the current helicity 
density evolution in cycle 23 
\citep[e.g.,][]{zetal10}.

We find that the standard error for  $\bar{\mathbf{A}}\cdot\bar{\mathbf{B}}$/R is less than 1 G$^2$ and has maximum at the sunspot formation latitudes. Similarly, the standard error of  $\mathbf{\overline{a\cdot\mathbf{b}}}/R$ is less than 100 G$^2$. \cite{2017SoPh292.29L} estimated the maximum uncertainty of the vector magnetic field   measurement about $\sim$ 10--20 G. Then, we can conclude that the standard deviation of  uncertainty of the determined the normalized magnetic vector potential components is less than 10 G. A more accurate estimate of errors for the magnetic helicity density on each synoptic map requires additional analysis, and it will be done in a separate paper.

\section{Discussion and Conclusions}
\label{sec:discussion}

Here, we propose a novel approach to reconstruct the surface 
magnetic helicity density on the Sun and sun-like stars. 
The magnetic vector potential is determined via 
decomposition of vector magnetic field in the
toroidal and poloidal components. The method is applied to study the evolution of magnetic helicity density of large- and small-scale fields in solar cycle 24. 

Our results show that at the beginning if cycle 24, 
the large- and small-scale magnetic helicities are opposite in sign to each other, and both show the hemispheric asymmetry in sign. This is in agreement with the magnetic helicity conservation theory \citep{pouquet-al:1975a}, which predicts the sign of variations of the  small-scale magnetic helicity density,  $\mathbf{\overline{a\cdot\mathbf{b}}}$ should be opposite to the sign of the large-scale helicity.
However, in the declining phase of cycle 24 the large-scale polar magnetic field shows the negative (positive) magnetic helicity in the North (South), which is the same in sign as the helicity of solar active regions (small-scale field).
Apparently, in the polar regions the large-scale helicity, $\mathbf{\overline{A}\cdot\mathbf{\overline{B}}}$,  changed the sign near the solar maximum. 
The dynamo models predict that both $\mathbf{\overline{A}\cdot\mathbf{\overline{B}}}$ and  $\mathbf{\overline{a\cdot\mathbf{b}}}$ can change sign  in course of the solar cycle due to other sources of the magnetic helicity evolution like the turbulent eddy diffusivity of the large-scale magnetic field \citep{kleruz82, pip13M,sok2013}. Another possible reason is the magnetic helicity flux escape from the solar surface \citep{2000JGR...10510481B}.
We also note that dynamo models of \citep{pip13M} predict another polar inversion of the large-scale magnetic helicity models during solar minima. Thus, 
we expect that the magnetic helicity of large-scale (polar field) will reverse its sign again around the solar minimum, which would restore the hemispheric 
helicity rule to the same condition as it was at the 
beginning of cycle 24.

The amplitude of magnetic helicity of large-scale fields is significantly (an order of magnitude) smaller as compared with helicity of small-scale (active region) fields. Thus, active regions 
appear to be the major contributors to magnetic 
 helicity observed on the visible solar surface (photosphere). This is further supported by the fact 
that the emergence of a single (large) active region 
could affect the hemispheric preference in sign of 
helicity for some Carrington rotations.

 In our data we see that the magnetic helicity density  of the large- and small-scale fields often show the same sign in each hemisphere. In our opinion this can be due to a bi-modal distribution of the large-scale toroidal magnetic field on the solar surface. Indeed, we see the two components of the toroidal magnetic field: one is near equator and another is in the mid latitudes. The near equatorial toroidal field is originated from sunspots. The origin of the mid-latitude toroidal field is unclear. This component of the large-scale field can results into mono-helical over the scales  magnetic field distributions in the each hemisphere. The evolution of the large-scale magnetic field and its vector potential results in a complicated magnetic helicity evolution. In the data, it is seen that the southern hemisphere was close to the bi-helical state during the growing phase of the solar cycle, from 2010 to 2012. The northern hemisphere, except the polar region, shows  the same sign of the magnetic helicity density for the large- and small-scale field during that period.    Another source of violation of the hemispheric helicity rule is the asymmetric  about equator development of  solar cycle 24. This results in the equatorial parity breaking in the large-scale magnetic field and vector potential components. In particular, the obtained evolution of $\bar{A}_{\phi}$ is different from previous results by \citet{black-bran:02} and \cite{PP14}.

The hemispheric helicity rule and bi-helical distributions of the solar magnetic fields are
essential properties of the solar dynamo operating in the convection zone. We find that in solar cycle 24  these properties of solar dynamo show a complicated evolution in the way which is not expected in any current dynamo model. The suggested method of the magnetic helicity reconstruction can be applied  to other stars where the low-degree modes of the vector magnetic field distributions, such as determined in \cite{2018MNRAS480477V}, can be used to calculate the magnetic vector potential and magnetic helicity density of the large-scale stellar magnetic field.

\acknowledgments
This work was borne out of discussions held among the authors during ``Solar Helicities in Theory 
and Observations: Implications for Space Weather and Dynamo Theory'' Program at Nordic Institute 
for Theoretical Physics (NORDITA) in 4--29 March 2019.
JSOC/HMImaps: Courtesy of NASA/SDO HMI science team. AAP acknowledges partial support by NASA 80NSSC17K0686 and NSF AGS-1620773 grants. AGK was partially supported by NASA grants NNX14AB70G and NNX17AE76A. 
VVP  conducted this work as a part of FR II.16 of ISTP SB RAS, also the support of the RFBR grant 19-52-53045 GFEN-a is acknowledged. 

\vspace{5mm}
\facilities{SDO/HMI}
\bibliographystyle{aasjournal}

\begin{thebibliography}{}
\expandafter\ifx\csname natexlab\endcsname\relax\def\natexlab#1{#1}\fi
\providecommand{\url}[1]{\href{#1}{#1}}
\providecommand{\dodoi}[1]{doi:~\href{http://doi.org/#1}{\nolinkurl{#1}}}
\providecommand{\doeprint}[1]{\href{http://ascl.net/#1}{\nolinkurl{http://ascl.net/#1}}}
\providecommand{\doarXiv}[1]{\href{https://arxiv.org/abs/#1}{\nolinkurl{https://arxiv.org/abs/#1}}}

\bibitem[{{Balasubramaniam} \& {Pevtsov}(2011)}]{Balasubramaniam.Pevtsov2011}
{Balasubramaniam}, K.~S., \& {Pevtsov}, A. 2011, in \procspie, Vol. 8148, Solar
  Physics and Space Weather Instrumentation IV, 814809

\bibitem[{{Bao} \& {Zhang}(1998)}]{baoZh98}
{Bao}, S., \& {Zhang}, H. 1998, \apj, 496, L43, \dodoi{10.1086/311232}

\bibitem[{{Berger} \& {Hornig}(2018)}]{2018JPhA51W5501B}
{Berger}, M.~A., \& {Hornig}, G. 2018, Journal of Physics A Mathematical
  General, 51, 495501, \dodoi{10.1088/1751-8121/aaea88}

\bibitem[{{Berger} \& {Ruzmaikin}(2000)}]{2000JGR...10510481B}
{Berger}, M.~A., \& {Ruzmaikin}, A. 2000, \jgr, 105, 10481,
  \dodoi{10.1029/1999JA900392}

\bibitem[{{Blackman} \& {Brandenburg}(2002)}]{black-bran:02}
{Blackman}, E.~G., \& {Brandenburg}, A. 2002, \apj, 579, 379,
  \dodoi{10.1086/342705}

\bibitem[{{Blackman} \& {Brandenburg}(2003)}]{bl-br2003}
---. 2003, \apjl, 584, L99, \dodoi{10.1086/368374}

\bibitem[{{Brandenburg} {et~al.}(2017){Brandenburg}, {Petrie}, \&
  {Singh}}]{2017ApJ836.21B}
{Brandenburg}, A., {Petrie}, G. J.~D., \& {Singh}, N.~K. 2017, \apj, 836, 21,
  \dodoi{10.3847/1538-4357/836/1/21}

\bibitem[{{Brandenburg} \& {Subramanian}(2005)}]{brsu05}
{Brandenburg}, A., \& {Subramanian}, K. 2005, \physrep, 417, 1,
  \dodoi{10.1016/j.physrep.2005.06.005}

\bibitem[{{Choudhuri} {et~al.}(2004){Choudhuri}, {Chatterjee}, \&
  {Nandy}}]{choud2004ApJ}
{Choudhuri}, A.~R., {Chatterjee}, P., \& {Nandy}, D. 2004, \apjl, 615, L57,
  \dodoi{10.1086/426054}

\bibitem[{Frisch {et~al.}(1975)Frisch, Pouquet, L\'eorat, \&
  A.}]{pouquet-al:1975a}
Frisch, U., Pouquet, A., L\'eorat, J., \& A., M. 1975, J. Fluid Mech., 68, 769

\bibitem[{Hagino \& Sakurai(2005)}]{hagsak05}
Hagino, M., \& Sakurai, T. 2005, Publ. Astron. Soc. Japan, 57, 481

\bibitem[{Kleeorin \& Ruzmaikin(1982)}]{kleruz82}
Kleeorin, N.~I., \& Ruzmaikin, A.~A. 1982, Magnetohydrodynamics, 18, 116

\bibitem[{Krause \& R\"adler(1980)}]{KR80}
Krause, F., \& R\"adler, K.-H. 1980, Mean-Field Magnetohydrodynamics and Dynamo
  Theory (Berlin: Akademie-Verlag), 271

\bibitem[{{Liu} {et~al.}(2017){Liu}, {Hoeksema}, {Sun}, \&
  {Hayashi}}]{2017SoPh292.29L}
{Liu}, Y., {Hoeksema}, J.~T., {Sun}, X., \& {Hayashi}, K. 2017, \solphys, 292,
  29, \dodoi{10.1007/s11207-017-1056-9}

\bibitem[{{Longcope} {et~al.}(1998){Longcope}, {Fisher}, \&
  {Pevtsov}}]{long1998ApJ}
{Longcope}, D.~W., {Fisher}, G.~H., \& {Pevtsov}, A.~A. 1998, \apj, 507, 417,
  \dodoi{10.1086/306312}

\bibitem[{{Pesnell} {et~al.}(2012){Pesnell}, {Thompson}, \&
  {Chamberlin}}]{Pesnell.etal2012}
{Pesnell}, W.~D., {Thompson}, B.~J., \& {Chamberlin}, P.~C. 2012, \solphys,
  275, 3, \dodoi{10.1007/s11207-011-9841-3}

\bibitem[{{Pevtsov} {et~al.}(2014){Pevtsov}, {Berger}, {Nindos}, {Norton}, \&
  {van Driel-Gesztelyi}}]{Pevtsov.etal2014}
{Pevtsov}, A.~A., {Berger}, M.~A., {Nindos}, A., {Norton}, A.~A., \& {van
  Driel-Gesztelyi}, L. 2014, \ssr, 186, 285, \dodoi{10.1007/s11214-014-0082-2}

\bibitem[{{Pevtsov} {et~al.}(1995){Pevtsov}, {Canfield}, \&
  {Metcalf}}]{pev95ApJ}
{Pevtsov}, A.~A., {Canfield}, R.~C., \& {Metcalf}, T.~R. 1995, \apjl, 440,
  L109, \dodoi{10.1086/187773}

\bibitem[{{Pevtsov} {et~al.}(2008){Pevtsov}, {Canfield}, {Sakurai}, \&
  {Hagino}}]{Pevtsov.etal2008}
{Pevtsov}, A.~A., {Canfield}, R.~C., {Sakurai}, T., \& {Hagino}, M. 2008, \apj,
  677, 719, \dodoi{10.1086/533435}

\bibitem[{{Pevtsov} \& {Latushko}(2000)}]{2000ApJ...528..999P}
{Pevtsov}, A.~A., \& {Latushko}, S.~M. 2000, \apj, 528, 999,
  \dodoi{10.1086/308227}

\bibitem[{Pipin(2018)}]{2dspy}
Pipin, V. 2018, VVpipin/2DSPDy 0.1.1, \dodoi{10.5281/zenodo.1413149}

\bibitem[{{Pipin} \& {Kosovichev}(2018)}]{2018ApJ867.145P}
{Pipin}, V.~V., \& {Kosovichev}, A.~G. 2018, \apj, 867, 145,
  \dodoi{10.3847/1538-4357/aae1fb}

\bibitem[{{Pipin} \& {Pevtsov}(2014)}]{PP14}
{Pipin}, V.~V., \& {Pevtsov}, A.~A. 2014, \apj, 789, 21,
  \dodoi{10.1088/0004-637X/789/1/21}

\bibitem[{{Pipin} {et~al.}(2013){Pipin}, {Zhang}, {Sokoloff}, {Kuzanyan}, \&
  {Gao}}]{pip13M}
{Pipin}, V.~V., {Zhang}, H., {Sokoloff}, D.~D., {Kuzanyan}, K.~M., \& {Gao}, Y.
  2013, \mnras, 435, 2581, \dodoi{10.1093/mnras/stt1465}

\bibitem[{Pouquet {et~al.}(1975)Pouquet, Frisch, \&
  L\'eorat}]{pouquet-al:1975b}
Pouquet, A., Frisch, U., \& L\'eorat, J. 1975, J. Fluid Mech., 68, 769

\bibitem[{{Raedler}(1986)}]{rad86AN}
{Raedler}, K.-H. 1986, Astronomische Nachrichten, 307, 89,
  \dodoi{10.1002/asna.2113070205}

\bibitem[{{Scherrer} {et~al.}(2012){Scherrer}, {Schou}, {Bush}, {Kosovichev},
  {Bogart}, {Hoeksema}, {Liu}, {Duvall}, {Zhao}, {Title}, {Schrijver},
  {Tarbell}, \& {Tomczyk}}]{Scherrer.etal2012}
{Scherrer}, P.~H., {Schou}, J., {Bush}, R.~I., {et~al.} 2012, \solphys, 275,
  207, \dodoi{10.1007/s11207-011-9834-2}

\bibitem[{{Seehafer}(1990)}]{see1990SoPh}
{Seehafer}, N. 1990, \solphys, 125, 219, \dodoi{10.1007/BF00158402}

\bibitem[{{Singh} {et~al.}(2018){Singh}, {K{\"a}pyl{\"a}}, {Brandenburg},
  {K{\"a}pyl{\"a}}, {Lagg}, \& {Virtanen}}]{2018arXiv180404994S}
{Singh}, N.~K., {K{\"a}pyl{\"a}}, M.~J., {Brandenburg}, A., {et~al.} 2018,
  \apj, 863, 182, \dodoi{10.3847/1538-4357/aad0f2}

\bibitem[{{Sokoloff} {et~al.}(2013){Sokoloff}, {Zhang}, {Moss}, {Kleeorin},
  {Kuzanyan}, {Rogachevski}, {Gao}, \& {Xu}}]{sok2013}
{Sokoloff}, D., {Zhang}, H., {Moss}, D., {et~al.} 2013, in Proceedings of the
  International Astronomical Union, Vol.~8, Symposium S294 (Solar and
  Astrophysical Dynamos and Magnetic Activity), ed. A.~G. {Kosovichev}, E.~{de
  Gouveia Dal Pino}, \& Y.~{Yan}, 313--318

\bibitem[{{Stenflo} \& {Guedel}(1988)}]{sten88}
{Stenflo}, J.~O., \& {Guedel}, M. 1988, \aap, 191, 137

\bibitem[{{Vidotto} {et~al.}(2018){Vidotto}, {Lehmann}, {Jardine}, \&
  {Pevtsov}}]{2018MNRAS480477V}
{Vidotto}, A.~A., {Lehmann}, L.~T., {Jardine}, M., \& {Pevtsov}, A.~A. 2018,
  \mnras, 480, 477, \dodoi{10.1093/mnras/sty1926}

\bibitem[{{Virtanen} {et~al.}(2019){Virtanen}, {Pevtsov}, \&
  {Mursula}}]{Virtanen.etal2019}
{Virtanen}, I.~I., {Pevtsov}, A.~A., \& {Mursula}, K. 2019, \aap, 624, A73,
  \dodoi{10.1051/0004-6361/201834895}

\bibitem[{{Wieczorek} \& {Meschede}(2018)}]{sht2018}
{Wieczorek}, M., \& {Meschede}, M. 2018, Geochemistry, Geophysics, Geosystems,
  19, 2574, \dodoi{10.1029/2018GC007529}

\bibitem[{{Zhang} {et~al.}(2010){Zhang}, {Sakurai}, {Pevtsov}, {Gao}, {Xu},
  {Sokoloff}, \& {Kuzanyan}}]{zetal10}
{Zhang}, H., {Sakurai}, T., {Pevtsov}, A., {et~al.} 2010, \mnras, 402, L30,
  \dodoi{10.1111/j.1745-3933.2009.00793.x}

\end{thebibliography}

\end{document}